\documentclass[sigconf]{acmart}

\usepackage{booktabs, multicol,multirow}
\usepackage{hyperref}
\usepackage{subfigure}
\usepackage{algorithm}
\usepackage{algorithmic}
\usepackage{makecell}
\usepackage{amsthm,amsmath}
\usepackage{mathrsfs}
\AtBeginDocument{%
  }

\copyrightyear{2024}
\acmYear{2024}
\setcopyright{acmlicensed}\acmConference[MM '24]{Proceedings of the 32nd ACM International Conference on Multimedia}{October 28-November 1, 2024}{Melbourne, VIC, Australia}
\acmBooktitle{Proceedings of the 32nd ACM International Conference on Multimedia (MM '24), October 28-November 1, 2024, Melbourne, VIC, Australia}
\acmDOI{10.1145/3664647.3688981}
\acmISBN{979-8-4007-0686-8/24/10}

\begin{document}

\title{Contrastive Learning-based Chaining-Cluster for Multilingual Voice-Face Association}

\author{Wuyang Chen}
\authornote{Both authors contributed equally to this research.}
\email{chenwuyangcn@gmail.com}
\author{Yanjie Sun}
\authornotemark[1]
\email{sunyanjie21@nudt.edu.cn}
\affiliation{%
  \institution{National University of Defense Technology}
  \city{Changsha}
  \country{China}
}

\author{Kele Xu}
\authornote{Corresponding author.}
\email{xukelele@163.com}
\affiliation{%
  \institution{National University of Defense Technology}
  \city{Changsha}
  \country{China}}

\author{Yong Dou}
\email{yongdou@nudt.edu.cn}
\affiliation{%
  \institution{National University of Defense Technology}
  \city{Changsha}
  \country{China}}

\renewcommand{\shortauthors}{Wuyang Chen, Yanjie Sun, Kele Xu and Yong Dou}

\begin{abstract}
The innate correlation between a person's face and voice has recently emerged as a compelling area of study, especially within the context of multilingual environments. This paper introduces our novel solution to the Face-Voice Association in Multilingual Environments (FAME) 2024 challenge, focusing on a contrastive learning-based chaining-cluster method to enhance face-voice association. This task involves the challenges of building biometric relations between auditory and visual modality cues and modelling the prosody interdependence between different languages while addressing both intrinsic and extrinsic variability present in the data. To handle these non-trivial challenges, our method employs supervised cross-contrastive (SCC) learning to establish robust associations between voices and faces in multi-language scenarios. Following this, we have specifically designed a chaining-cluster-based post-processing step to mitigate the impact of outliers often found in unconstrained in the wild data. We conducted extensive experiments to investigate the impact of language on face-voice association. The overall results were evaluated on the FAME public evaluation platform, where we achieved 2nd place. The results demonstrate the superior performance of our method, and we validate the robustness and effectiveness of our proposed approach. Code is available at \url{https://github.com/colaudiolab/FAME24_solution}.
\end{abstract}




\maketitle

\section{Introduction}
Human beings possess the ability to associate voices with faces, even for unknown individuals~\cite{mavica2013matching,belin2004thinking,kamachi2003putting}.
  In recent studies, deep neural networks can use computational methods to mimic humans in associating unheard voices with unseen face images~\cite{wen2021seeking,ijcai2022p526,morgado2021audio}.
 This capability can aid investigative work and biometric authentication, propel the advancement of virtual reality technologies, and even help anthropologists explore the physiological correlation between voiceprints and facial features, showcasing its usefulness in practical security fields and its academic value in research~\cite{kim2019learning,nagrani2018learnable,nagrani2018seeing}. 

 However, sustainable efforts~\cite{wen2021seeking,ijcai2022p526,morgado2021audio,DBLP:conf/icassp/SaeedNKZNYM23} show impressive performance on voice-face associating tasks, these works only focus on a single language environment, assuming that each person speaks only one language. With the growing prevalence of multilingual speakers and the complex dynamics of polyglot interactions, the original assumption can easily be invalidated, as half of the population is bilingual~\cite{mathews2019half}. When the same person speaks more than one language, we still expect the model to robustness match the correct facial image with different language speakings. However, the multilingual scenario makes the original task even more challenging. These challenges manifest in the following ways: In a monolingual setting, the network needs to learn the implicit mapping relationship between facial images and voices, which may involve understanding biometric relationships between prominent facial attributes and acoustic features. For instance, a person with thick eyebrows may have a deeper voice, while a person with high cheekbones may have a louder voice~\cite{wells2013perceptions,penton2001symmetry}; In a multilingual setting, the network needs to further identify similarities in the same person when speaking different languages. However, different languages have distinct vowels, consonants, grammar rules, and so on~\cite{mok2023similar}, making it hard to find similarities. This may involve uncovering hidden prosody, such as tone and stress patterns~\cite{van2021interplay}.

This article will focus on the association of voice-face under multilingual condition.
In this field,~\cite{nawaz2021cross} investigated the effects of multiple languages on voice-face association and speaker recognition tasks.
For the association task, they used two sub-networks to extract embeddings from faces and voices, followed by two fully connected layers. The final outputs from the two branches were optimized by a hinge loss to reduce the modality gap between them. Based on this architecture, they found that language does indeed have an effect on the voice-face association task, and a similar conclusion was drawn for speaker recognition. Further, their team organized a competition called the Face-voice Association in Multilingual Environments (FAME) Challenge 2024~\cite{saeed2024face}. The challenge used publicly available MAV-Celeb dataset for this task, which includes audio-visual data of 154 celebrities speaking three languages (English, Hindi, and Urdu). Along with this dataset, they also employ Fusion and Orthogonal Projection (FOP)~\cite{Saeed_2022} as their baseline. The architecture of the FOP method is similar to the one used in~\cite{nawaz2021cross}, but with improvements including attention-based fusion and the imposition of orthogonality constraints on the optimization objective. The MAV-Celeb and other monolingual voice-face datasets are collected from the Internet using an automatic pipeline, which inevitably contains the outliers. For example, face images or voice segments with the same identity label may not come from the same person, or there may be other voices or objects covering the target. It is important to note that the term ``outliers'' used here refers to both samples with incorrect labels and those with noisy or non-target interference. These outliers would cause confusion for the association method, especially during the inference stage, and can easily lead to incorrect matching results. However, these outliers are rarely taken into account in prior works. 

This work proposes a contrastive learning-based chaining-cluster method to handle mentioned challenges.
Firstly, considering the significant modality gap between voice and face image features, we design a two-branch network to align these multimodal representations in the same space with the help of supervised cross-contrastive (SCC) loss. The positive samples defined in the contrastive objective are voice-face pairs that belong to the same person, regardless of the speaking language. 
This definition of positive samples implicitly encourages the network to automatically learn the bilingual similarities when the same person speaks different languages.
Secondly, we obtain the initial test pair scores by feeding pairs into the SCC network and computing a normalized similarity on the output representations. These representations are then used with a chaining-cluster on both modalities. Initially, they are divided into male and female clusters, and then for each gender under each modality, samples closer to the nearest cluster center are finely clustered into identities. For high-confidence identity cluster results (closer to the cluster center, called prototypes), we calculate the cross-modality prototype similarity. The final prototype similarity results, as well as the gender cluster results, are used to refine the initial test scores. The refined scores can greatly eliminate the impact of outliers, which is achieved by using high-confidence cluster results (inliers) to guide outliers.

Our proposed method can achieve high performance on voice-face association tasks for both heard and unheard languages, demonstrating the effectiveness of the pipeline. 
Overall, our contributions include the following aspects:
\begin{itemize}
    \item We proposed a novel framework integrating contrastive learning with a chaining-cluster approach to enhance voice-face association in multilingual settings.
    \item We designed a multimodal representation learning model based on supervised cross-contrastive learning, which effectively aligns voice and face features in a shared embedding space, facilitating robust association even in the presence of multilingual data. 
    \item We developed a chaining-cluster post-processing method that handles outliers and improves the accuracy of the association task by leveraging high-confidence cluster results to refine initial test scores.
    \item We achieved second place in the 2024 FAME competition, demonstrating the effectiveness and robustness of our proposed method in real-world, multilingual scenarios.
\end{itemize}
\begin{figure*}
     \centering
         \includegraphics[width=0.94\textwidth]{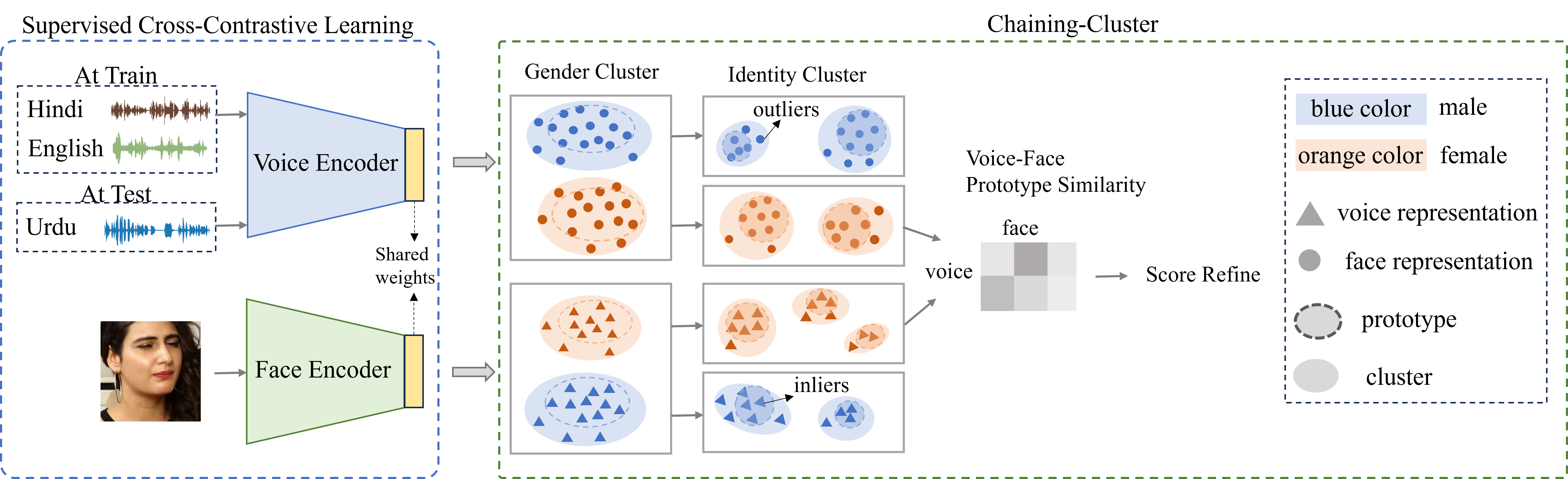}
         \caption{Our method consists of a supervised cross-contrastive learning phase and a chaining-cluster phase. The first phase accepts a pair of voice-face and outputs their similarity score (We use V1 unheard as an illustrative example). The chaining-cluster phase is employed only during the testing phase, as a post-processing step to refine the initial score from the first phase.}
    
        \label{fig:method}
      
\end{figure*}
\section{Related Work}
The burgeoning intersection of auditory and visual data analysis has catalyzed significant advancements in the field of audio-visual association tasks. Particularly within the context of speaker recognition ~\cite{tao2023speaker,shah2023speaker,fan2020cn,jabeen2024multimodal,chung2018voxceleb2} and biometric identification ~\cite{DBLP:conf/interspeech/MarkitantovRR022,folorunso2019review,kusmierczyk2020biometric}, the ability to accurately link voices with corresponding facial images is paramount. However, the complexity of this task escalates in multilingual environments, where audio-visual to the multilingual speaker further challenges traditional models when only one of the languages has been heard. This section delves into the existing body of work that has paved the way for our current investigation, Including the cutting-edge research in face-voice association and the domain of multilingual speaker recognition.
\subsection{Face-Voice Association}
The field of face-voice association has seen significant progress in recent years, with researchers exploring innovative methods to understand and model the relationship between a person's face and their voice. Some studies have trained unimodal features separately and established face and voice association by learning joint embeddings~\cite{kim2018learning,wen2021seeking,nawaz2021cross,Saeed_2022}.
Saeed et al.~\cite{Saeed_2022} introduced a method known as Fusion and Orthogonal Projection (FOP) to improve face-voice association. They hypothesized that an enriched feature representation and efficient supervision were necessary for a discriminative joint embedding space. 
Other works have proposed methods involving cross-modal prototype contrastive learning and self-supervised learning approaches that align multimodal data during the feature learning phase~\cite{Chen_2024,Zhang_2021,Stevenage_2024}.
Chen et al.~\cite{Chen_2024} explored the generation of faces from voice segments with their method VoiceStyle, which combined cross-modal representation learning with generation modeling. They introduced cross-modal prototype contrastive learning to establish voice-face associations and used StyleGAN to produce faces guided by the learned alignment.
Additionally, self-supervised approaches gained traction. Zhang et al.~\cite{Zhang_2021} proposed self-supervised curriculum learning (SSCL) to enhance audio-visual association by exploiting the concurrency property as a latent supervision signal. Stevenage et al.~\cite{Stevenage_2024} indicates that the correlation between face and voice processing is stronger for familiar individuals, suggesting the importance of mental representation and neural cross-talk in processing familiar stimuli.

The method we propose aligns multimodalities during the representation learning phase. The purpose of doing this is to ensure that the features of different modalities are consistent in the feature space, thereby better achieving the association task of faces and voices.
\subsection{Multilingual Speaker Recognition}
Research on multilingual speech association has made some progress in several speech research fields. These include voice conversion~\cite{huu2022proposing}, speaker verification~\cite{sano2023cross} and recognition~\cite{woldemariam2020adapting}, language factors~\cite{crisostomi2022play}, speech recognition and synthesis, and multilingual speech recognition and speech recognition~\cite {saeed2020cross, athish2023multilingual}.
In work~\cite{liao2022robustness}, they first attempted to explore the feasibility of the original speaker recognition system in a cross-lingual context. They collected speech data from the same speaker speaking with Mandarin, English, and Taiwanese languages, forming a total of 40 speakers. And they tried different network structures to validate the system's recognition performance. The results indicated that in a cross-lingual scenario, the system could still maintain close to 80\% accuracy, although the performance decreased significantly compared to a single language model. This preliminary evidence suggests that a system can recognize a person regardless of the language they are speaking. 
However, it should be noted that the data collected in the work~\cite{liao2022robustness} was gathered in a relatively stable environment. Participants were asked to read given text prompts in different languages and answer five specified questions in their preferred language. The data collected in this experimental setting contains less noise and disturbance compared to what may exist in real-world scenarios, potentially leading to distribution differences.
Subsequently, Yik Heng et al.~\cite{ng2024multi} designed a multilingual (Malay, English, Mandarin, and Tamil languages) speaker verification system for a door security system in the real-world scenario. This work primarily focuses on the verification accuracy of the system without discussing the impact on system performance when language changes occur in the testing scenario.
From the previous work, it can be observed that their data collection environment is relatively homogeneous, which may not accurately represent the distribution of in-the-wild data. On the other hand, Nawaz et al. collected multilingual videos (English, Urdu and Hindi) from online video platforms, including political debates, press conferences, and indoor and outdoor interviews. The audio in these videos encompasses various complex scenarios that may exist in the real world, such as background conversations, music, and speech overlap. This dataset is known as MAV-Celeb ~\cite{nawaz2021cross}. They conduct both voice-face association and speaker recognition tasks to answer the question of whether these tasks are language-dependent.
\section{Methodology}
This paper proposes a contrastive learning-based chaining-cluster method for the Face-Voice Association in Multilingual Environment (FAME) 2024 challenge. Our approach begins by utilizing Supervised cross-contrastive (SCC) learning to establish associations between voices and faces (as detailed in subsection~\ref{subsec:SCC}). Subsequently, the learned face and voice representations are employed in a post-processing phase that incorporates the chaining-cluster score refinement technique, with the aim of addressing outliers prevalent in wild data (as elaborated in subsection~\ref{subsec:CC}). The overall schematic of the method is illustrated in Figure.\ref{fig:method}. Considering that there are four different test scenarios in FAME (V1: trained on Urdu and tested on Urdu; trained on Urdu and tested on English; trained on English and tested on Urdu; trained on English and tested on English), for clarity, we will use one scenario (trained on Urdu and tested on Urdu) as an example.
\subsection{Supervised Cross-Contrastive Learning}
\label{subsec:SCC}
To enhance the understanding of the association between voice and facial features of the same identity, we propose a novel network architecture grounded in SCC learning. This architecture comprises two distinct branches, denoted as Voice Encoder $E_v$ and Face Encoder $E_f$, each containing 11 Transformer layers. Additionally, a single Transformer layer with shared weights is positioned between the voice and face branches. This shared-weight layer improves learning efficiency and fosters the model's ability to recognize common patterns across voice and facial data.
We utilize the following loss function:
\begin{equation}
    \mathcal{L} = -\frac{1}{N}\sum_{i=1}^{N}\left[\log\frac{exp(E_v(v_i)^T E_f(f_i)/\tau)}{\sum_{j=1}^{N}exp(E_v(v_i)^T E_f(f_j)/\tau)}\right]
\end{equation}
where $N$ is the batch size. $\tau$ is the temperature hyper-parameter. $v_i$ and $f_i$ represent the voice and face that belong to the same identity, while the denominator is the sum of all samples in the current batch.
The trained SCC network is used to extract representations from test pairs. The initial score of the test pair $(v_i,f_i)$ is calculated using $1-cos(E_v(v_i),E_f(f_i))$. Where $cos(\cdot)$ is a normalized cosine similarity operation. 
\begin{algorithm}
\caption{Algorithm of Chaining-cluster post-processing.}
\begin{flushleft}
\hspace*{\algorithmicindent} \textbf{Input} test pairs $TEST$, initial $score$,\\
$\qquad \qquad$ voice, face representations $\{V,F\}$ from SCC \\
 \hspace*{\algorithmicindent} \textbf{Output} refined $score$
 \end{flushleft}
\begin{algorithmic}[1]

\FOR{ each modality $m$ in $\{V, F\}$} 
\STATE cluster $m$ into Male and Female clusters $S^m=\{S^m_\mathcal{M}, S^m_\mathcal{F}\}$
\FOR{ each sample $i$ in $S^m$}
\IF{ distance of $i$ to gender cluster center > $T^m$  }
\STATE $\widehat{S^m} \leftarrow  S^m \setminus  \{ i \}$
\ENDIF
\ENDFOR

\FOR{ cluster $S^{m,g}$ in $\{\widehat{S^m_\mathcal{M}}, \widehat{S^m_\mathcal{F}} \}$ }  
\STATE cluster $S^{m,g}$ into identity clusters $S^{m,g}=\{S^{m,g}_1,...,S^{m,g}_n\}$

\FOR{ each sample $j$ in $S^{m,g}$}
\IF{ distance of $j$ to identity cluster center > $T^g$  }

\STATE $\widehat{S^{m,g}} \leftarrow  S^{m,g} \setminus  \{ j \}$
\ENDIF
\ENDFOR
\STATE compute prototype $p^{m,g}$ for clusters in $\widehat{S^{m,g}}$
\ENDFOR
\ENDFOR

\FOR{ each gender $g$ in $\{\mathcal{M}, \mathcal{F} \}$ }
\STATE $sim^g \leftarrow$ compute\ Similarity($p^{m=V,g},p^{m=F,g}$)
\FOR{ $(v,f)$ in $TEST$}
\IF{$(v,f) \in S^{m,g}$}
\STATE $score_{v,f} = score_{v,f} \downarrow \text{ if } sim^g(v,f) > T^{\alpha}$
\ENDIF
\ENDFOR
\ENDFOR

\FOR{ $(v,f)$ in $TEST$}
\IF{$(v,f) \in S^m$}
\STATE  $C_v =$ clusterCenter($v$), $C_f =$ clusterCenter($f$)

\STATE $score_{v,f} = score_{v,f} \uparrow  \text{ if }  C_v \neq C_f $
\ENDIF
\ENDFOR
\end{algorithmic}
\end{algorithm}
\subsection{Chaining-Cluster}
\label{subsec:CC}
The rationale for implementing clustering is based on the fact that voice-face pairs are collected in uncontrolled environments, leading to numerous potential distractions. For audio data, these distractions may include background music, overlapping voices from other individuals, and environmental noise. For image data, distractions might involve varying facial orientations, makeup, and lighting conditions. Directly applying our SCC model to derive cross-modal similarity scores can result in inaccuracies due to these distractions. Clustering offers a robust solution by effectively managing minor distractions. For example, if an audio clip of a target male individual contains overlapping voices from females, the clustering process can still yield accurate results. This is because the primary component of the audio segment remains the male voice, with the female voices contributing minimally.

Our chaining-cluster pipeline consists of four primary steps, as delineated in Algorithm 1. These steps will be elaborated upon subsequently in this section. For illustrative purposes, we will use the SCC model trained on Urdu and corresponding pairs from the Urdu test set as examples throughout this discussion.

\textbf{Gender Cluster.}
For the voice-face test pairs, we initially cluster each modality into two clusters, intended to represent male ($\mathcal{M}$) and female ($\mathcal{F}$) categories. For the audio modality, we utilize the SCC model obtained from the first stage to extract audio representations $V = \{E_v(v_i)\}_{i=1}^{|\text{TEST}|}$ using its voice branch.
Subsequently, we apply the K-Means algorithm to these representations, clustering them into two categories.
Similarly, for the image modality, we extract face representations $F = \{E_f(f_i)\}_{i=1}^{|\text{TEST}|}$ using the SCC face branch and apply K-Means for clustering. For each modality, we then calculate the $L_2$ distance of each sample $i$ to its respective cluster center. If this distance exceeds the predefined threshold $T^{m}$, we remove those samples from the current candidate set $S^{m}$, thereby forming a high-confidence set $\widehat{S^{m}}$.

\textbf{Identity Cluster. }
After obtaining the gender cluster candidates for each modality, we aim further to cluster identity in each modality and gender candidate set $S^{m,g}$. For instance, consider the face modality from the female candidate set $\widehat{S^{m}_\mathcal{F}}$. We apply K-Means to test samples exist in  $\widehat{S^{m}_\mathcal{F}}$ into $n$ groups. The number of clusters $n$, is determined by both the Elbow method and test statistics. 

\begin{figure}
     \centering
         \includegraphics[width=0.35\textwidth]{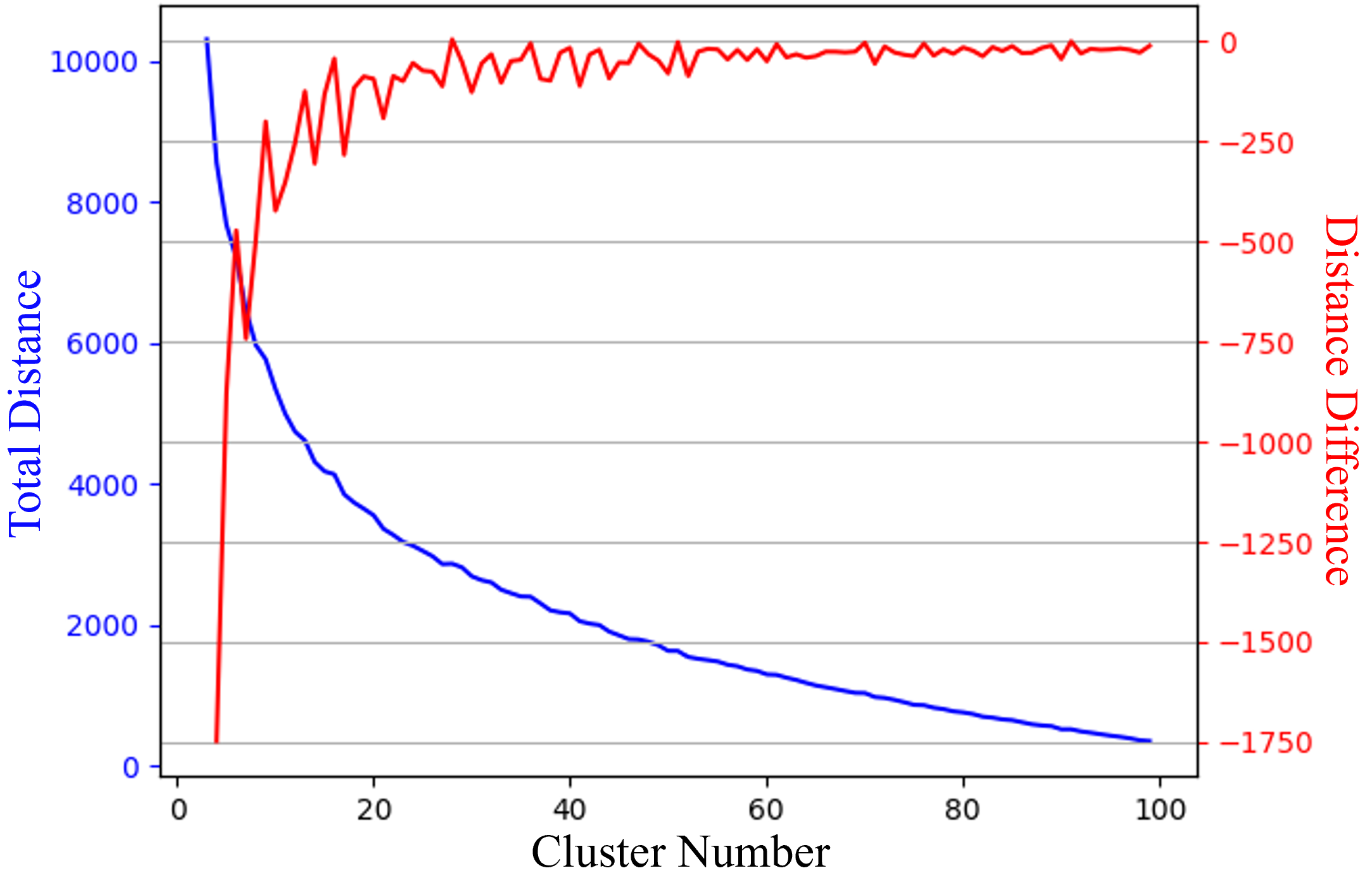}
         \caption{The blue line represents the total $L_2$ distance of each sample to its cluster center, and the red line represents the distance difference between adjacent cluster numbers.}
    
        \label{fig:elbow}
\end{figure}

The Elbow method uses the total sum of $L_2$ distances of each sample to its cluster center as an indicator to find the optimal number of clusters. The clustering results for $\widehat{S^{m}_\mathcal{F}}$ are shown in Figure \ref{fig:elbow}. We select the optimal cluster number $n$ by considering the index of the large difference value that is closer to half the total number of test pairs (we halve the count because we assume the numbers of male and female tests are nearly equal). In our experiment, we set $n=16$ for heard Urdu test pairs.

\begin{figure}
     \centering
         \includegraphics[width=0.28\textwidth]{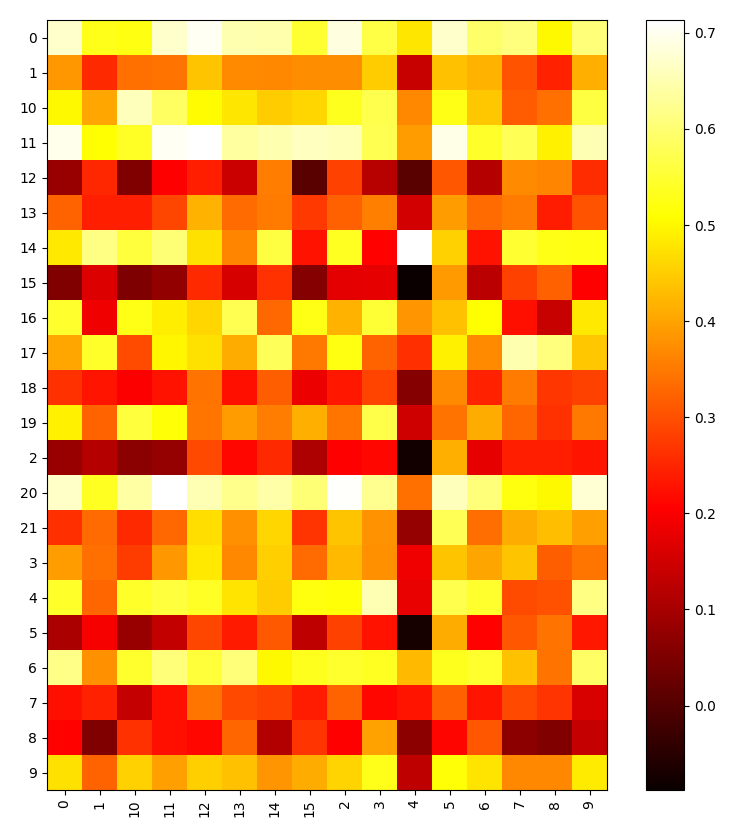}
         \caption{Cross-modal prototype similarity matrix. The raw represents voice prototypes, and the column represents the face prototypes.}
    
        \label{fig:similarity}
\end{figure}

\textbf{Voice-Face Prototype Similarity. }Next, we propose a prototype-based cross-modal similarity metric. We believe that outliers can be identified in the SCC representation space by measuring the distance of each sample to its nearest cluster center. The specific approach is as follows: First, the prototypes $p^{m,g}$ are calculated by averaging the representations in the high-confidence cluster set of $\widehat{S^{m,g}}$. The high-confidence cluster set is defined by the distance to its nearest identity cluster center, which is lower than a threshold. This means that the prototypes exclude the outliers. The similarity matrix is then computed through the normalized cosine similarity between voice prototypes and face prototypes.
\begin{equation}
    Sim(p^{m=V,g},p^{m=F,g}) = \frac{E_v({p^V_i})\cdot E_f({p^F_j})}{||E_v({p^V_i})||\cdot|| E_f({p^F_j})||}
    \label{eq:similarity}
\end{equation}
where $p^V_i$ and $p^F_j$ represent the corresponding averaging voice and face representations from high confidence sets. The similarity matrix of these two modal prototypes, $sim^{g=\mathcal{F}}$ when the gender is chosen as female, is visualized as shown in Figure \ref{fig:similarity}. The brighter color indicates a higher similarity between the corresponding voice and face prototypes, suggesting a high probability that the voice-face test pairs belonging to these highly similar prototypes are from the same person. This assumption is used as a principle to refine our initial test score.

\textbf{Score Refine. }
Our main intention in this step is to 1) use the cross-modal prototype similarity obtained from inliers to adjust the initial scores of both inliers and outlier samples. When calculating the similarity score between voice and face on individual samples, it is inevitable that the score will be affected by sample noise and differences. However, clustering can largely mitigate the impact of individual differences. To make the results of similarity calculations more robust, we exclude outliers from computing the cross-modal prototype similarity. If satisfy some principles, the scores of inliers and outlier samples will be refined. 2) use the gender cluster results to adjust the initial scores. Specifically, we propose score refine method based on two principles:
\begin{itemize}
   \item    Penalize pairs of voice and face gender mismatches based on gender clusters.
   \item    Reward pairs with high cross-modal prototype similarity based on identity clusters. 
\end{itemize}
First, We compute the upper bound $B_u(s)$ and lower bound $B_l(s)$ of score base on the maximum and minimum value of initial score. Then, we perform a gender mismatch penalty: 
\begin{equation}
    score(v_i,f_i) = \left\{
    \begin{array}{lcl}
        B_u(s)          &        & if \ C_{v_i}\neq C_{f_i}\\
        score(v_i,f_i)  &        & else
    \end{array}
    \right.
\label{eq:punish}
\end{equation}
If a test pair \((v_i, f_i)\) is found in the set \({S^m}\), we determine its nearest cluster centers, denoted as \(C_v\) and \(C_f\). A gender mismatch is identified if \(C_v \neq C_f\). In such cases, the upper bound is employed to adjust the score accordingly.

Finally, we perform cross-modal prototype high similarity rewarding using the following formula: 
\begin{equation}
    score(v_i,f_i) =\left \{
    \begin{array}{ll}
     B_l(s) + [score(v_i,f_i)-B_l(s)]*\alpha     & if \ sim^g(v,f) < T^\alpha \\
     socre(v_i,f_i)                              & else 
    \end{array}
    \right.
\label{eq:award}
\end{equation}
For a given test pair \((v_i, f_i)\) that exists in \(S^{m,g}\), if the similarity between the corresponding voice and face prototype, denoted as \(sim^g(v,f)\), is lower than the predefined threshold \(T^{\alpha}\), the distance fraction is reduced as a reward. Here, \(\alpha\) serves as a refinement factor.
\section{Experiments and Results}
The system evaluation is determined by the output scores for each test pair. The score for each pair represents the system's confidence in the identity match between the voice segment and face image. Our work utilizes the public MAV-Celeb dataset to train and test. The test process is taken by the competition-provided platform. 
\begin{table}[b]
\caption{The details of the MAV-Celeb dataset.}
\centering
\begin{tabular}{lccc}
\hline
{\bf Dataset} & {\bf E/U/V        $1$        -EU} & {\bf E/H/V        $2$        -EH} \\ 
\hline
\hline
 \#  of Celebrities & 70 & 84  \\ 
\hline
 \#  of male celebrities & 40/40/\textbf{3} & 53/53/\textbf{3}  \\ 
\hline
 \#  of female celebrities & 24/24/\textbf{3} & 25/25/\textbf{3}  \\ 
\hline
 \#  of videos & 402/555/957 & 646/484/1130  \\ 
\hline
 \#  of hours & 30/54/84 & 51/33/84  \\ 
\hline
Avg  \#  of videos/celebrity & 6/8/14 & 8/6/14  \\ 
\hline
\end{tabular}
\label{tab:data_stats}
\end{table}
\subsection{Datasets}
The MAV-Celeb is an audio-visual dataset containing multilingual speaking videos. The dataset covers a total of 154 identities, focusing on the English, Hindi, and Urdu languages. The videos are collected from online media platforms with an automatic collection pipeline. Specifically, for each person's name from the list of persons of interest (POIs), the English, Hindi, and Urdu version videos are downloaded and post-processed, forming the MAV-Celeb dataset. The videos are all real-world and cover various scenes, such as political debates, outdoor/indoor interviews, conferences, and even public movie clips, containing unconstrained noise degradation, which poses challenges to the task.

To better analyze the language effect on the voice-face association task, the dataset is split into two versions. The first version (V1) contains English and Urdu corpora, while the second version (V2) includes English and Hindi languages. The details of the dataset are illustrated in Table \ref{tab:data_stats}. Both versions have 6 identities in the test set. 
\subsection{Evaluation Metric}
The evaluation metric used by the testing platform is the equal error rate (EER). This metric calculates the false acceptance rate (FAR) and false rejection rate (FRR) based on multiple predefined thresholds. When a threshold is found that equalizes the FAR and FRR, it represents the final EER. In reality, a system may encounter various situations where it cannot ensure that the language of the test voice matches that of the training set. To comprehensively assess the system's robustness, the evaluation encompasses two scenarios. One involves the test language being included in the training data, while the other scenario assumes that the test language does not appear in the training dataset at all. Thus, the final output comprises a total of four items, and the overall evaluation score is based on the average of these EERs.
\subsection{Implementation Details}
For the supervised cross-contrastive learning phase, we first pretrain the model on the public Voxceleb1~\cite{nagrani2017voxceleb} dataset. It should be noted that the pretraining step does not start from random weights; instead, it uses parameters that have been trained on the VGGSound~\cite{chen2020vggsound} and AudioSet~\cite{gemmeke2017audio} datasets. During fine-tuning, we use both the training data from the V1 and V2 datasets, as well as English, Hindi, and Urdu interview videos crawled from the Internet (processed using the same pipeline as mentioned in ~\cite{nawaz2021cross}). It should be noted that for the unheard test scenario, we exclude the testing language from the training set.
\subsection{Evaluation Results}
The competition is ultimately evaluated on the V1 dataset, and the details of the overall score on V1 are shown in Table \ref{result}. Our performance significantly outperforms the baseline, achieving second place compared to all other participating teams. We also provide the results of our system on the V2 dataset for reference, which are displayed in Table \ref{result2}. It should be noted that the method we used to evaluate V2 does not include any post-processing. It only utilizes the SCC network along with the baseline's linear fusion architecture. The results on V2 also show significant improvement compared to the baseline, indicating that the SCC network has learned powerful discriminative features that are important for the voice-face association task.
\begin{table}
	\caption{\textbf{The evaluation results.} V1 }
	\label{result}
	\centering
 \small
  \renewcommand\arraystretch{0.95}
	\begin{tabular}{*{5}{c}} 
		\toprule
		 Method & Config & Eng. test & Urdu test & Overall score \\  \midrule
            \multirow{2}{*}{Ours}  & Eng. train & 17.1 & 28.2 & \multirow{2}{*}{20.5}    \\
            & Urdu train & 18.3 & 18.4 & \\
         
		\bottomrule
	\end{tabular}
\end{table}
\begin{table}
	\caption{\textbf{The evaluation results.} V2 }
	\label{result2}
	\centering
 \small
  \renewcommand\arraystretch{0.95}
	\begin{tabular}{*{5}{c}} 
		\toprule
		 Method & Config & Eng. test & Hindi test & Overall score \\  \midrule
		  \multirow{2}{*}{FOP~\cite{Saeed_2022}}  & Eng. train & 20.8 & 24.0 & \multirow{2}{*}{22.0}    \\
            & Hindi train & 24.0 & 19.3 & \\
           
        \midrule
            \multirow{2}{*}{Ours}  & Eng. train & 18.9 & 19.9 & \multirow{2}{*}{20.4}    \\
            & Hindi train & 22.1 & 20.7 & \\
         
		\bottomrule
	\end{tabular}
\end{table}
\subsection{Ablation Study}
To quantitatively measure how much the post-processing proposed in this paper improves overall performance, we removed the post-processing step and calculated the scores on the V1 dataset using only the representations extracted from the SCC network. The results are presented in Table \ref{abaltion}. It can be observed that the post-processing effectively reduces the EER, especially in cases where the original EER is relatively high. These findings also indicate that the chaining-cluster post-processing we proposed can effectively utilize high-confidence samples (inliers) to guide the outliers.
\begin{table}
	\caption{\textbf{The ablation results.} V1 }
	\label{abaltion}
	\centering
 \small
  \renewcommand\arraystretch{0.95}
	\begin{tabular}{*{5}{c}} 
		\toprule
		 Method & Config & Eng. test & Urdu test & Overall score \\  \midrule
		 

         \multirow{2}{*}{\makecell[c]{w/o \\ Score Refine}}  & Eng. train & 32.6 & 34.3 & \multirow{2}{*}{29.6}    \\
            & Urdu train & 25.2 & 26.1 & \\
           
		\bottomrule
	\end{tabular}
\end{table}
\section{Discussion and Conclusions}
\subsection{Discussion}
We believe that the success of our system firstly stems from the supervised cross-contrastive learning.  Existing research indicates that native speakers can easily distinguish non-native accents from second-language speakers, and prosody remains correlated even when the same individual speaks different languages.  This prosody may encompass various aspects such as rhythm, stress, and even intonation patterns.  However, what these relevant factors currently entail is still a black box.  Our training process relies on supervised face-voice pairing with different languages.  This type of training pair forces the model to associate the facial image of the same person with different languages while distinguishing it from others.  Secondly, the post-processing further enhances our system.  The chaining-cluster method gradually removes outliers and uses the remaining inliers to adjust the initial scores.  This approach ensures that performance does not degrade in noisy real-world scenarios. In upcoming projects, we will investigate using memory-based methods for more elegant end-to-end prediction.
\subsection{Conclusions}
In this study, we addressed the challenge of voice-face association in multilingual contexts by proposing a novel solution that utilizes the SCC network combined with chaining-cluster-based post-processing. This approach effectively addresses the difficulties associated with cross-modal and cross-language associations, as well as the outlier influences commonly encountered in real-world data. As a result, our method demonstrated significant improvements over the baseline, achieving promising scores and robustness in identifying matching pairs in both heard and unheard scenarios.
\begin{acks}
This work is supported by the National Science and Technology Major Project (2023ZD0121101).
\end{acks}

\clearpage
\newpage


\balance
\bibliographystyle{ACM-Reference-Format}

\begin{thebibliography}{41}


\ifx \showCODEN    \undefined \def \showCODEN     #1{\unskip}     \fi
\ifx \showDOI      \undefined \def \showDOI       #1{#1}\fi
\ifx \showISBNx    \undefined \def \showISBNx     #1{\unskip}     \fi
\ifx \showISBNxiii \undefined \def \showISBNxiii  #1{\unskip}     \fi
\ifx \showISSN     \undefined \def \showISSN      #1{\unskip}     \fi
\ifx \showLCCN     \undefined \def \showLCCN      #1{\unskip}     \fi
\ifx \shownote     \undefined \def \shownote      #1{#1}          \fi
\ifx \showarticletitle \undefined \def \showarticletitle #1{#1}   \fi
\ifx \showURL      \undefined \def \showURL       {\relax}        \fi
\providecommand\bibfield[2]{#2}
\providecommand\bibinfo[2]{#2}
\providecommand\natexlab[1]{#1}
\providecommand\showeprint[2][]{arXiv:#2}

\bibitem[Athish et~al\mbox{.}(2023)]%
        {athish2023multilingual}
\bibfield{author}{\bibinfo{person}{A.~Y. Athish}, \bibinfo{person}{S.~K G}, {and} \bibinfo{person}{Sivakumar M}.} \bibinfo{year}{2023}\natexlab{}.
\newblock \showarticletitle{Multilingual Speech Recognition Using Reinforcement Learning}.
\newblock \bibinfo{journal}{\emph{International Conference on Computing Communication and Networking Technologies}} (\bibinfo{year}{2023}).
\newblock
\urldef\tempurl%
\url{dblp.org/rec/conf/icccnt/AthishGM23}
\showURL{%
\tempurl}


\bibitem[Belin et~al\mbox{.}(2004)]%
        {belin2004thinking}
\bibfield{author}{\bibinfo{person}{Pascal Belin}, \bibinfo{person}{Shirley Fecteau}, {and} \bibinfo{person}{Catherine Bedard}.} \bibinfo{year}{2004}\natexlab{}.
\newblock \showarticletitle{Thinking the voice: neural correlates of voice perception}.
\newblock \bibinfo{journal}{\emph{Trends in cognitive sciences}} \bibinfo{volume}{8}, \bibinfo{number}{3} (\bibinfo{year}{2004}), \bibinfo{pages}{129--135}.
\newblock


\bibitem[Chen et~al\mbox{.}(2020)]%
        {chen2020vggsound}
\bibfield{author}{\bibinfo{person}{Honglie Chen}, \bibinfo{person}{Weidi Xie}, \bibinfo{person}{Andrea Vedaldi}, {and} \bibinfo{person}{Andrew Zisserman}.} \bibinfo{year}{2020}\natexlab{}.
\newblock \showarticletitle{Vggsound: A large-scale audio-visual dataset}. In \bibinfo{booktitle}{\emph{ICASSP 2020-2020 IEEE International Conference on Acoustics, Speech and Signal Processing (ICASSP)}}. IEEE, \bibinfo{pages}{721--725}.
\newblock


\bibitem[Chen et~al\mbox{.}(2024)]%
        {Chen_2024}
\bibfield{author}{\bibinfo{person}{Wuyang Chen}, \bibinfo{person}{Boqing Zhu}, \bibinfo{person}{Kele Xu}, \bibinfo{person}{Yong Dou}, {and} \bibinfo{person}{Dawei Feng}.} \bibinfo{year}{2024}\natexlab{}.
\newblock \showarticletitle{VoiceStyle: Voice-based Face Generation Via Cross-modal Prototype Contrastive Learning}.
\newblock \bibinfo{journal}{\emph{ACM Transactions on Multimedia Computing, Communications, and Applications (TOMCCAP)}} (\bibinfo{year}{2024}).
\newblock
\urldef\tempurl%
\url{https://doi.org/10.1145/3671002}
\showDOI{\tempurl}


\bibitem[Chung et~al\mbox{.}(2018)]%
        {chung2018voxceleb2}
\bibfield{author}{\bibinfo{person}{Joon~Son Chung}, \bibinfo{person}{Arsha Nagrani}, {and} \bibinfo{person}{Andrew Zisserman}.} \bibinfo{year}{2018}\natexlab{}.
\newblock \showarticletitle{VoxCeleb2: Deep Speaker Recognition}. In \bibinfo{booktitle}{\emph{19th Annual Conference of the International Speech Communication Association, Interspeech 2018, Hyderabad, India, September 2-6, 2018}}, \bibfield{editor}{\bibinfo{person}{B.~Yegnanarayana}} (Ed.). \bibinfo{publisher}{{ISCA}}, \bibinfo{pages}{1086--1090}.
\newblock


\bibitem[Crisostomi et~al\mbox{.}(2022)]%
        {crisostomi2022play}
\bibfield{author}{\bibinfo{person}{Donato Crisostomi}, \bibinfo{person}{Davide Bernardi}, {and} \bibinfo{person}{Sarah Campbell}.} \bibinfo{year}{2022}\natexlab{}.
\newblock \showarticletitle{Play música alegre: A Large-Scale Empirical Analysis of Cross-Lingual Phenomena in Voice Assistant Interactions}.
\newblock \bibinfo{journal}{\emph{MMNLU}} (\bibinfo{year}{2022}).
\newblock


\bibitem[Fan et~al\mbox{.}(2020)]%
        {fan2020cn}
\bibfield{author}{\bibinfo{person}{Yue Fan}, \bibinfo{person}{JW Kang}, \bibinfo{person}{LT Li}, \bibinfo{person}{KC Li}, \bibinfo{person}{HL Chen}, \bibinfo{person}{ST Cheng}, \bibinfo{person}{PY Zhang}, \bibinfo{person}{ZY Zhou}, \bibinfo{person}{YQ Cai}, {and} \bibinfo{person}{Dong Wang}.} \bibinfo{year}{2020}\natexlab{}.
\newblock \showarticletitle{Cn-celeb: a challenging chinese speaker recognition dataset}. In \bibinfo{booktitle}{\emph{ICASSP 2020-2020 IEEE International Conference on Acoustics, Speech and Signal Processing (ICASSP)}}. IEEE, \bibinfo{pages}{7604--7608}.
\newblock


\bibitem[Folorunso et~al\mbox{.}(2019)]%
        {folorunso2019review}
\bibfield{author}{\bibinfo{person}{CO Folorunso}, \bibinfo{person}{OS Asaolu}, {and} \bibinfo{person}{OP Popoola}.} \bibinfo{year}{2019}\natexlab{}.
\newblock \showarticletitle{A review of voice-base person identification: state-of-the-art}.
\newblock \bibinfo{journal}{\emph{Covenant Journal of Engineering Technology}} (\bibinfo{year}{2019}).
\newblock


\bibitem[Gemmeke et~al\mbox{.}(2017)]%
        {gemmeke2017audio}
\bibfield{author}{\bibinfo{person}{Jort~F Gemmeke}, \bibinfo{person}{Daniel~PW Ellis}, \bibinfo{person}{Dylan Freedman}, \bibinfo{person}{Aren Jansen}, \bibinfo{person}{Wade Lawrence}, \bibinfo{person}{R~Channing Moore}, \bibinfo{person}{Manoj Plakal}, {and} \bibinfo{person}{Marvin Ritter}.} \bibinfo{year}{2017}\natexlab{}.
\newblock \showarticletitle{Audio set: An ontology and human-labeled dataset for audio events}. In \bibinfo{booktitle}{\emph{2017 IEEE international conference on acoustics, speech and signal processing (ICASSP)}}. IEEE, \bibinfo{pages}{776--780}.
\newblock


\bibitem[Huu et~al\mbox{.}(2022)]%
        {huu2022proposing}
\bibfield{author}{\bibinfo{person}{P.~N. Huu}, \bibinfo{person}{Sy-Tuyen Ho}, \bibinfo{person}{Chau Nguyen~Le Bao}, \bibinfo{person}{M. Anh}, \bibinfo{person}{Luong~Nguyen Thien}, \bibinfo{person}{Hieu~Nguyen Duc}, \bibinfo{person}{Minh-Trien Pham}, \bibinfo{person}{Dat~Vu Tien}, {and} \bibinfo{person}{Q. Minh}.} \bibinfo{year}{2022}\natexlab{}.
\newblock \showarticletitle{Proposing XLSR Multilingual Model for Vietnamese Language Recognition}.
\newblock \bibinfo{journal}{\emph{Conference on Research, Innovation and Vision for the Future in Computing \& Communication Technologies}} (\bibinfo{year}{2022}).
\newblock
\urldef\tempurl%
\url{dblp.org/rec/conf/rivf/HuuHBATDPTM22}
\showURL{%
\tempurl}


\bibitem[Jabeen et~al\mbox{.}(2024)]%
        {jabeen2024multimodal}
\bibfield{author}{\bibinfo{person}{Summaira Jabeen}, \bibinfo{person}{Muhammad~Shoib Amin}, {and} \bibinfo{person}{Xi Li}.} \bibinfo{year}{2024}\natexlab{}.
\newblock \showarticletitle{Multimodal pre-train then transfer learning approach for speaker recognition}.
\newblock \bibinfo{journal}{\emph{Multimedia Tools and Applications}} (\bibinfo{year}{2024}), \bibinfo{pages}{1--14}.
\newblock


\bibitem[Kamachi et~al\mbox{.}(2003)]%
        {kamachi2003putting}
\bibfield{author}{\bibinfo{person}{Miyuki Kamachi}, \bibinfo{person}{Harold Hill}, \bibinfo{person}{Karen Lander}, {and} \bibinfo{person}{Eric Vatikiotis-Bateson}.} \bibinfo{year}{2003}\natexlab{}.
\newblock \showarticletitle{Putting the face to the voice': Matching identity across modality}.
\newblock \bibinfo{journal}{\emph{Current Biology}} \bibinfo{volume}{13}, \bibinfo{number}{19} (\bibinfo{year}{2003}), \bibinfo{pages}{1709--1714}.
\newblock


\bibitem[Kim et~al\mbox{.}(2018)]%
        {kim2018learning}
\bibfield{author}{\bibinfo{person}{Changil Kim}, \bibinfo{person}{Hijung~Valentina Shin}, \bibinfo{person}{Tae-Hyun Oh}, \bibinfo{person}{Alexandre Kaspar}, \bibinfo{person}{Mohamed Elgharib}, {and} \bibinfo{person}{Wojciech Matusik}.} \bibinfo{year}{2018}\natexlab{}.
\newblock \showarticletitle{On learning associations of faces and voices}. In \bibinfo{booktitle}{\emph{Asian Conference on Computer Vision}}. Springer, \bibinfo{pages}{276--292}.
\newblock


\bibitem[Kim et~al\mbox{.}(2019)]%
        {kim2019learning}
\bibfield{author}{\bibinfo{person}{Changil Kim}, \bibinfo{person}{Hijung~Valentina Shin}, \bibinfo{person}{Tae-Hyun Oh}, \bibinfo{person}{Alexandre Kaspar}, \bibinfo{person}{Mohamed Elgharib}, {and} \bibinfo{person}{Wojciech Matusik}.} \bibinfo{year}{2019}\natexlab{}.
\newblock \showarticletitle{On learning associations of faces and voices}. In \bibinfo{booktitle}{\emph{Computer Vision--ACCV 2018: 14th Asian Conference on Computer Vision, Perth, Australia, December 2--6, 2018, Revised Selected Papers, Part V 14}}. Springer, \bibinfo{pages}{276--292}.
\newblock


\bibitem[Ku{\'s}mierczyk et~al\mbox{.}(2020)]%
        {kusmierczyk2020biometric}
\bibfield{author}{\bibinfo{person}{Aleksander Ku{\'s}mierczyk}, \bibinfo{person}{Martyna S{\l}awi{\'n}ska}, \bibinfo{person}{Kornel {\.Z}aba}, {and} \bibinfo{person}{Khalid Saeed}.} \bibinfo{year}{2020}\natexlab{}.
\newblock \showarticletitle{Biometric fusion system using face and voice recognition: a comparison approach: biometric fusion system using face and voice characteristics}.
\newblock \bibinfo{journal}{\emph{Advanced Computing and Systems for Security: Volume Ten}} (\bibinfo{year}{2020}), \bibinfo{pages}{71--89}.
\newblock


\bibitem[Liao et~al\mbox{.}(2022)]%
        {liao2022robustness}
\bibfield{author}{\bibinfo{person}{Wen-Hung Liao}, \bibinfo{person}{Wei-Yu Chen}, {and} \bibinfo{person}{Yi-Chieh Wu}.} \bibinfo{year}{2022}\natexlab{}.
\newblock \showarticletitle{On the robustness of cross-lingual speaker recognition using transformer-based approaches}. In \bibinfo{booktitle}{\emph{2022 26th International Conference on Pattern Recognition (ICPR)}}. IEEE, \bibinfo{pages}{366--371}.
\newblock


\bibitem[Markitantov et~al\mbox{.}(2022)]%
        {DBLP:conf/interspeech/MarkitantovRR022}
\bibfield{author}{\bibinfo{person}{Maxim Markitantov}, \bibinfo{person}{Elena Ryumina}, \bibinfo{person}{Dmitry Ryumin}, {and} \bibinfo{person}{Alexey Karpov}.} \bibinfo{year}{2022}\natexlab{}.
\newblock \showarticletitle{Biometric Russian Audio-Visual Extended {MASKS} {(BRAVE-MASKS)} Corpus: Multimodal Mask Type Recognition Task}. In \bibinfo{booktitle}{\emph{23rd Annual Conference of the International Speech Communication Association, Interspeech 2022, Incheon, Korea, September 18-22, 2022}}, \bibfield{editor}{\bibinfo{person}{Hanseok Ko} {and} \bibinfo{person}{John H.~L. Hansen}} (Eds.). \bibinfo{publisher}{{ISCA}}, \bibinfo{pages}{1756--1760}.
\newblock
\urldef\tempurl%
\url{https://doi.org/10.21437/INTERSPEECH.2022-10240}
\showDOI{\tempurl}


\bibitem[Mathews(2019)]%
        {mathews2019half}
\bibfield{author}{\bibinfo{person}{Jay Mathews}.} \bibinfo{year}{2019}\natexlab{}.
\newblock \showarticletitle{Half of the world is bilingual. What’s our problem}.
\newblock \bibinfo{journal}{\emph{The Washington Post}}  \bibinfo{volume}{25} (\bibinfo{year}{2019}).
\newblock


\bibitem[Mavica and Barenholtz(2013)]%
        {mavica2013matching}
\bibfield{author}{\bibinfo{person}{Lauren~W Mavica} {and} \bibinfo{person}{Elan Barenholtz}.} \bibinfo{year}{2013}\natexlab{}.
\newblock \showarticletitle{Matching voice and face identity from static images.}
\newblock \bibinfo{journal}{\emph{Journal of Experimental Psychology: Human Perception and Performance}} \bibinfo{volume}{39}, \bibinfo{number}{2} (\bibinfo{year}{2013}), \bibinfo{pages}{307}.
\newblock


\bibitem[Mok et~al\mbox{.}(2023)]%
        {mok2023similar}
\bibfield{author}{\bibinfo{person}{Peggy~PK Mok}, \bibinfo{person}{Holly~SH Fung}, \bibinfo{person}{Grace~WL Cao}, {and} \bibinfo{person}{Chun~Wai Leung}.} \bibinfo{year}{2023}\natexlab{}.
\newblock \showarticletitle{How similar are the formants in the speech of bilingual speakers?}
\newblock \bibinfo{journal}{\emph{International Journal of Speech, Language \& the Law}} \bibinfo{volume}{30}, \bibinfo{number}{1} (\bibinfo{year}{2023}).
\newblock


\bibitem[Morgado et~al\mbox{.}(2021)]%
        {morgado2021audio}
\bibfield{author}{\bibinfo{person}{Pedro Morgado}, \bibinfo{person}{Nuno Vasconcelos}, {and} \bibinfo{person}{Ishan Misra}.} \bibinfo{year}{2021}\natexlab{}.
\newblock \showarticletitle{Audio-visual instance discrimination with cross-modal agreement}. In \bibinfo{booktitle}{\emph{Proceedings of the IEEE/CVF conference on computer vision and pattern recognition}}. \bibinfo{pages}{12475--12486}.
\newblock


\bibitem[Nagrani et~al\mbox{.}(2018a)]%
        {nagrani2018learnable}
\bibfield{author}{\bibinfo{person}{Arsha Nagrani}, \bibinfo{person}{Samuel Albanie}, {and} \bibinfo{person}{Andrew Zisserman}.} \bibinfo{year}{2018}\natexlab{a}.
\newblock \showarticletitle{Learnable pins: Cross-modal embeddings for person identity}. In \bibinfo{booktitle}{\emph{Proceedings of the European conference on computer vision (ECCV)}}. \bibinfo{pages}{71--88}.
\newblock


\bibitem[Nagrani et~al\mbox{.}(2018b)]%
        {nagrani2018seeing}
\bibfield{author}{\bibinfo{person}{Arsha Nagrani}, \bibinfo{person}{Samuel Albanie}, {and} \bibinfo{person}{Andrew Zisserman}.} \bibinfo{year}{2018}\natexlab{b}.
\newblock \showarticletitle{Seeing voices and hearing faces: Cross-modal biometric matching}. In \bibinfo{booktitle}{\emph{Proceedings of the IEEE conference on computer vision and pattern recognition}}. \bibinfo{pages}{8427--8436}.
\newblock


\bibitem[Nagrani et~al\mbox{.}(2017)]%
        {nagrani2017voxceleb}
\bibfield{author}{\bibinfo{person}{Arsha Nagrani}, \bibinfo{person}{Joon~Son Chung}, {and} \bibinfo{person}{Andrew Zisserman}.} \bibinfo{year}{2017}\natexlab{}.
\newblock \showarticletitle{Voxceleb: a large-scale speaker identification dataset}.
\newblock \bibinfo{journal}{\emph{arXiv preprint arXiv:1706.08612}} (\bibinfo{year}{2017}).
\newblock


\bibitem[Nawaz et~al\mbox{.}(2021)]%
        {nawaz2021cross}
\bibfield{author}{\bibinfo{person}{Shah Nawaz}, \bibinfo{person}{Muhammad~Saad Saeed}, \bibinfo{person}{Pietro Morerio}, \bibinfo{person}{Arif Mahmood}, \bibinfo{person}{Ignazio Gallo}, \bibinfo{person}{Muhammad~Haroon Yousaf}, {and} \bibinfo{person}{Alessio Del~Bue}.} \bibinfo{year}{2021}\natexlab{}.
\newblock \showarticletitle{Cross-modal speaker verification and recognition: A multilingual perspective}. In \bibinfo{booktitle}{\emph{Proceedings of the IEEE/CVF conference on computer vision and pattern recognition}}. \bibinfo{pages}{1682--1691}.
\newblock


\bibitem[Ng and Hong(2024)]%
        {ng2024multi}
\bibfield{author}{\bibinfo{person}{Yik~Heng Ng} {and} \bibinfo{person}{Kai~Sze Hong}.} \bibinfo{year}{2024}\natexlab{}.
\newblock \showarticletitle{Multi-Lingual Speaker Verification Using Malay, English, Mandarin and Tamil Languages for Door Security System}. In \bibinfo{booktitle}{\emph{2024 3rd International Conference on Digital Transformation and Applications (ICDXA)}}. IEEE, \bibinfo{pages}{109--114}.
\newblock


\bibitem[Penton-Voak et~al\mbox{.}(2001)]%
        {penton2001symmetry}
\bibfield{author}{\bibinfo{person}{Ian~S Penton-Voak}, \bibinfo{person}{Benedict~C Jones}, \bibinfo{person}{Anthony~C Little}, \bibinfo{person}{S Baker}, \bibinfo{person}{Burt Tiddeman}, \bibinfo{person}{DM Burt}, {and} \bibinfo{person}{David~I Perrett}.} \bibinfo{year}{2001}\natexlab{}.
\newblock \showarticletitle{Symmetry, sexual dimorphism in facial proportions and male facial attractiveness}.
\newblock \bibinfo{journal}{\emph{Proceedings of the Royal Society of London. Series B: Biological Sciences}} \bibinfo{volume}{268}, \bibinfo{number}{1476} (\bibinfo{year}{2001}), \bibinfo{pages}{1617--1623}.
\newblock


\bibitem[Saeed et~al\mbox{.}(2022)]%
        {Saeed_2022}
\bibfield{author}{\bibinfo{person}{Muhammad~Saad Saeed}, \bibinfo{person}{Muhammad~Haris Khan}, \bibinfo{person}{Shah Nawaz}, \bibinfo{person}{Muhammad~Haroon Yousaf}, {and} \bibinfo{person}{Alessio~Del Bue}.} \bibinfo{year}{2022}\natexlab{}.
\newblock \showarticletitle{Fusion and Orthogonal Projection for Improved Face-Voice Association}. In \bibinfo{booktitle}{\emph{{IEEE} International Conference on Acoustics, Speech and Signal Processing, {ICASSP} 2022, Virtual and Singapore, 23-27 May 2022}}. \bibinfo{publisher}{{IEEE}}, \bibinfo{pages}{7057--7061}.
\newblock
\urldef\tempurl%
\url{https://doi.org/10.1109/ICASSP43922.2022.9747704}
\showDOI{\tempurl}


\bibitem[Saeed et~al\mbox{.}(2023)]%
        {DBLP:conf/icassp/SaeedNKZNYM23}
\bibfield{author}{\bibinfo{person}{Muhammad~Saad Saeed}, \bibinfo{person}{Shah Nawaz}, \bibinfo{person}{Muhammad~Haris Khan}, \bibinfo{person}{Muhammad~Zaigham Zaheer}, \bibinfo{person}{Karthik Nandakumar}, \bibinfo{person}{Muhammad~Haroon Yousaf}, {and} \bibinfo{person}{Arif Mahmood}.} \bibinfo{year}{2023}\natexlab{}.
\newblock \showarticletitle{Single-branch Network for Multimodal Training}. In \bibinfo{booktitle}{\emph{{IEEE} International Conference on Acoustics, Speech and Signal Processing {ICASSP} 2023, Rhodes Island, Greece, June 4-10, 2023}}. \bibinfo{publisher}{{IEEE}}, \bibinfo{pages}{1--5}.
\newblock
\urldef\tempurl%
\url{https://doi.org/10.1109/ICASSP49357.2023.10097207}
\showDOI{\tempurl}


\bibitem[Saeed et~al\mbox{.}(2020)]%
        {saeed2020cross}
\bibfield{author}{\bibinfo{person}{M.~S. Saeed}, \bibinfo{person}{Shah Nawaz}, \bibinfo{person}{Pietro Morerio}, \bibinfo{person}{A. Mahmood}, \bibinfo{person}{I. Gallo}, \bibinfo{person}{M. Yousaf}, {and} \bibinfo{person}{A.~D. Bue}.} \bibinfo{year}{2020}\natexlab{}.
\newblock \showarticletitle{Cross-modal Speaker Verification and Recognition: A Multilingual Perspective}.
\newblock \bibinfo{journal}{\emph{2021 IEEE/CVF Conference on Computer Vision and Pattern Recognition Workshops (CVPRW)}} (\bibinfo{year}{2020}).
\newblock
\urldef\tempurl%
\url{dblp.org/rec/journals/corr/abs-2004-13780}
\showURL{%
\tempurl}


\bibitem[Saeed et~al\mbox{.}(2024)]%
        {saeed2024face}
\bibfield{author}{\bibinfo{person}{Muhammad~Saad Saeed}, \bibinfo{person}{Shah Nawaz}, \bibinfo{person}{Muhammad~Salman Tahir}, \bibinfo{person}{Rohan~Kumar Das}, \bibinfo{person}{Muhammad~Zaigham Zaheer}, \bibinfo{person}{Marta Moscati}, \bibinfo{person}{Markus Schedl}, \bibinfo{person}{Muhammad~Haris Khan}, \bibinfo{person}{Karthik Nandakumar}, {and} \bibinfo{person}{Muhammad~Haroon Yousaf}.} \bibinfo{year}{2024}\natexlab{}.
\newblock \showarticletitle{Face-voice Association in Multilingual Environments (FAME) Challenge 2024 Evaluation Plan}.
\newblock \bibinfo{journal}{\emph{arXiv preprint arXiv:2404.09342}} (\bibinfo{year}{2024}).
\newblock


\bibitem[Sano et~al\mbox{.}(2023)]%
        {sano2023cross}
\bibfield{author}{\bibinfo{person}{Ryotaro Sano}, \bibinfo{person}{Masafumi Nishida}, \bibinfo{person}{Satoru Tsuge}, \bibinfo{person}{Shingo Kuroiwa}, {and} \bibinfo{person}{Hiroyuki Yoshimura}.} \bibinfo{year}{2023}\natexlab{}.
\newblock \showarticletitle{Cross-Lingual Speaker Identification for Japanese-English Bilinguals}.
\newblock \bibinfo{journal}{\emph{Global Conference on Consumer Electronics}} (\bibinfo{year}{2023}).
\newblock
\urldef\tempurl%
\url{dblp.org/rec/conf/gcce/SanoNTKY23}
\showURL{%
\tempurl}


\bibitem[Shah et~al\mbox{.}(2023)]%
        {shah2023speaker}
\bibfield{author}{\bibinfo{person}{Saqlain~Hussain Shah}, \bibinfo{person}{Muhammad~Saad Saeed}, \bibinfo{person}{Shah Nawaz}, {and} \bibinfo{person}{Muhammad~Haroon Yousaf}.} \bibinfo{year}{2023}\natexlab{}.
\newblock \showarticletitle{Speaker recognition in realistic scenario using multimodal data}. In \bibinfo{booktitle}{\emph{2023 3rd International Conference on Artificial Intelligence (ICAI)}}. IEEE, \bibinfo{pages}{209--213}.
\newblock


\bibitem[Stevenage et~al\mbox{.}(2024)]%
        {Stevenage_2024}
\bibfield{author}{\bibinfo{person}{S. Stevenage}, \bibinfo{person}{Rebecca Edey}, \bibinfo{person}{Rebecca Keay}, \bibinfo{person}{Rebecca Morrison}, {and} \bibinfo{person}{David~J. Robertson}.} \bibinfo{year}{2024}\natexlab{}.
\newblock \showarticletitle{Familiarity Is Key: Exploring the Effect of Familiarity on the Face-Voice Correlation}.
\newblock \bibinfo{journal}{\emph{Brain Science}} (\bibinfo{year}{2024}).
\newblock
\urldef\tempurl%
\url{https://doi.org/10.3390/brainsci14020112}
\showDOI{\tempurl}


\bibitem[Tao et~al\mbox{.}(2023)]%
        {tao2023speaker}
\bibfield{author}{\bibinfo{person}{Ruijie Tao}, \bibinfo{person}{Kong~Aik Lee}, \bibinfo{person}{Zhan Shi}, {and} \bibinfo{person}{Haizhou Li}.} \bibinfo{year}{2023}\natexlab{}.
\newblock \showarticletitle{Speaker recognition with two-step multi-modal deep cleansing}. In \bibinfo{booktitle}{\emph{ICASSP 2023-2023 IEEE International Conference on Acoustics, Speech and Signal Processing (ICASSP)}}. IEEE, \bibinfo{pages}{1--5}.
\newblock


\bibitem[Van~Maastricht et~al\mbox{.}(2021)]%
        {van2021interplay}
\bibfield{author}{\bibinfo{person}{Lieke Van~Maastricht}, \bibinfo{person}{Tim Zee}, \bibinfo{person}{Emiel Krahmer}, {and} \bibinfo{person}{Marc Swerts}.} \bibinfo{year}{2021}\natexlab{}.
\newblock \showarticletitle{The interplay of prosodic cues in the L2: How intonation, rhythm, and speech rate in speech by Spanish learners of Dutch contribute to L1 Dutch perceptions of accentedness and comprehensibility}.
\newblock \bibinfo{journal}{\emph{Speech Communication}}  \bibinfo{volume}{133} (\bibinfo{year}{2021}), \bibinfo{pages}{81--90}.
\newblock


\bibitem[Wells et~al\mbox{.}(2013)]%
        {wells2013perceptions}
\bibfield{author}{\bibinfo{person}{Timothy Wells}, \bibinfo{person}{Thom Baguley}, \bibinfo{person}{Mark Sergeant}, {and} \bibinfo{person}{Andrew Dunn}.} \bibinfo{year}{2013}\natexlab{}.
\newblock \showarticletitle{Perceptions of human attractiveness comprising face and voice cues}.
\newblock \bibinfo{journal}{\emph{Archives of sexual behavior}}  \bibinfo{volume}{42} (\bibinfo{year}{2013}), \bibinfo{pages}{805--811}.
\newblock


\bibitem[Wen et~al\mbox{.}(2021)]%
        {wen2021seeking}
\bibfield{author}{\bibinfo{person}{Peisong Wen}, \bibinfo{person}{Qianqian Xu}, \bibinfo{person}{Yangbangyan Jiang}, \bibinfo{person}{Zhiyong Yang}, \bibinfo{person}{Yuan He}, {and} \bibinfo{person}{Qingming Huang}.} \bibinfo{year}{2021}\natexlab{}.
\newblock \showarticletitle{Seeking the shape of sound: An adaptive framework for learning voice-face association}. In \bibinfo{booktitle}{\emph{Proceedings of the IEEE/CVF conference on computer vision and pattern recognition}}. \bibinfo{pages}{16347--16356}.
\newblock


\bibitem[Woldemariam and Dahlgren(2020)]%
        {woldemariam2020adapting}
\bibfield{author}{\bibinfo{person}{Y. Woldemariam} {and} \bibinfo{person}{Adam Dahlgren}.} \bibinfo{year}{2020}\natexlab{}.
\newblock \showarticletitle{Adapting Language Specific Components of Cross-Media Analysis Frameworks to Less-Resourced Languages: the Case of Amharic}.
\newblock \bibinfo{journal}{\emph{Workshop on Spoken Language Technologies for Under-resourced Languages}} (\bibinfo{year}{2020}).
\newblock
\urldef\tempurl%
\url{dblp.org/rec/conf/sltu/WoldemariamD20}
\showURL{%
\tempurl}


\bibitem[Zhang et~al\mbox{.}(2021)]%
        {Zhang_2021}
\bibfield{author}{\bibinfo{person}{Jingran Zhang}, \bibinfo{person}{Xing Xu}, \bibinfo{person}{Fumin Shen}, \bibinfo{person}{Huimin Lu}, \bibinfo{person}{Xin Liu}, {and} \bibinfo{person}{Heng~Tao Shen}.} \bibinfo{year}{2021}\natexlab{}.
\newblock \showarticletitle{Enhancing Audio-Visual Association with Self-Supervised Curriculum Learning}.
\newblock \bibinfo{journal}{\emph{AAAI Conference on Artificial Intelligence}} (\bibinfo{year}{2021}).
\newblock
\urldef\tempurl%
\url{https://doi.org/10.1609/aaai.v35i4.16447}
\showDOI{\tempurl}


\bibitem[Zhu et~al\mbox{.}(2022)]%
        {ijcai2022p526}
\bibfield{author}{\bibinfo{person}{Boqing Zhu}, \bibinfo{person}{Kele Xu}, \bibinfo{person}{Changjian Wang}, \bibinfo{person}{Zheng Qin}, \bibinfo{person}{Tao Sun}, \bibinfo{person}{Huaimin Wang}, {and} \bibinfo{person}{Yuxing Peng}.} \bibinfo{year}{2022}\natexlab{}.
\newblock \showarticletitle{Unsupervised Voice-Face Representation Learning by Cross-Modal Prototype Contrast}. In \bibinfo{booktitle}{\emph{Proceedings of the Thirty-First International Joint Conference on Artificial Intelligence, {IJCAI-22}}}, \bibfield{editor}{\bibinfo{person}{Lud~De Raedt}} (Ed.). \bibinfo{publisher}{International Joint Conferences on Artificial Intelligence Organization}, \bibinfo{pages}{3787--3794}.
\newblock
\urldef\tempurl%
\url{https://doi.org/10.24963/ijcai.2022/526}
\showDOI{\tempurl}
\newblock
\shownote{Main Track}.


\end{thebibliography}

\end{document}